# Dirac-vortex modes beyond the continuum limit

Jiayu Fan[1,2], Jiusi Yu[1], Aoning Luo[1], Yiyi Yao[1], Shijie Kang[1], Xiexuan Zhang[1], Haitao Li[1,*], Biye Xie[3], Xiao-Dong Chen[4], Xiaoxiao Wu[1,*]

[1]Modern Matter Laboratory and Advanced Materials Thrust, The Hong Kong University of Science and Technology (Guangzhou), Nansha, Guangzhou 511400, Guangdong, China

[2]Department of Electronic and Computer Engineering, The Hong Kong University of Science and Technology, Clear Water Bay, Kowloon, Hong Kong, China

[3]School of Science and Engineering, The Chinese University of Hong Kong, Shenzhen, Guangdong 518172, P.R. China.

[4]School of Physics & State Key Laboratory of Optoelectronic Materials and Technologies, Sun Yat-sen University, Guangzhou 510275, China

[*]Corresponding authors: H. Li (haitaoli@hkust-gz.edu.cn), X. Wu (xwuan@connect.ust.hk)

## Abstract

Dirac-vortex modes (DVMs) in Kekulé-modulated lattices provide a topological route to wave confinement and are commonly described by the continuum Jackiw–Rossi model, in which the initial phase acts as a redundant gauge degree of freedom and does not affect observables of the mode. Here we show that this picture breaks down in discrete lattices when the complex mass texture that induces the DVMs no longer satisfies the slowly varying envelope approximation. In this regime, lattice discreteness turns the initial phase into a physically observable parameter that shifts the DVM center. By further introducing a sublattice-antisymmetric perturbation, we convert this phase-dependent center motion into a continuous spectral response of the DVM, enabling its frequency tuning across nearly the entire topological bandgap. Our simulation and experimental results agree well with a revised continuum model accounting for the mode-center motion. Within this perturbative framework, the model shows that the frequency shift exhibits a sinusoidal-like dependence on the initial phase. These findings reveal initial phase-sensitivity of the DVMs realized in lattices, an important and basic feature absent from the ideal continuum Jackiw–Rossi model, and demonstrate initial phase engineering as a potential pathway towards reconfigurable photonic devices.

## Introduction

Topological defect states provide a powerful route for confining light and sound through band topology rather than conventional local defects [1–4]. Among them, Dirac-vortex modes (DVMs) in Kekulé-modulated lattices have emerged as a versatile platform [5–9] for practical applications such as topological cavities [10–14] and fiber waveguides [15–19], owing to their strong localization, spectral isolation, guaranteed single-mode properties and scalable mode volumes [20–23].

To date, research and engineering on these DVM states have predominantly relied on the ideal continuum model proposed by Jackiw and Rossi due to the direct mapping between the Hamiltonian and the Kekulé-modulated lattice [24–26]. In this mapping, the complex Kekulé-order parameter in the model is mapped to the amplitude and phase of the aperiodic Kekulé distortion of the lattice. Although the model provides a powerful theoretical framework in the continuum limit, it inherently smooths out discreteness of the lattice. This approximation raises a critical question: does this model capture the full physical picture of DVMs in actual discrete realizations? Specifically, the extent to which the intrinsic discreteness can generate observable consequences absent from the continuum modeling remains elusive. Resolving this issue would determine whether the discrete nature can unlock hidden degrees of freedom for controlling lattice-based DVMs.

Here, we show that the DVMs in a discrete lattice exhibit a feature qualitatively absent from the ideal continuum model, especially when the vortex modulation approaches a constant amplitude profile. Merely a redundant gauge degree of freedom (DOF) in the ideal continuum model, the global initial phase of the Kekulé modulation becomes physically consequential in a discrete lattice and drives motion of the mode center. By introducing a sublattice-antisymmetric perturbation across a mirror axis, we convert this initial phase-dependent motion into a continuous spectral response of the DVM. As the initial phase varies a complete $2\pi$ period, the DVM traverses around the vortex core with opposite sublattice biases, achieving continuous and precise frequency tuning across nearly the entire topological bandgap. A perturbative approach in the continuum framework analytically links the frequency shift and the initial phase through the mode-center motion, which agrees well with numerical and experimental results. More broadly, by treating the discrete lattice as an active design space, this lattice-based paradigm expands topological-photonic design and offers a transferable framework for exploring Dirac-vortex physics across acoustic, elastic-wave, and other wave platforms.

## Results

We start from a quasi-2D photonic crystal (PhC) and focus on TM modes [Fig. 1(a)]. For demonstration purposes, we choose specific geometric parameters, resulting in an effective permittivity of $\varepsilon_{\mathrm{eff}} = 8.3$ for the dielectric rods in the quasi-2D approximation [27]. Through a band-folding scheme, a four-fold Dirac point emerges at 8.85 GHz (referred to below as the Dirac frequency) at Γ point (see SI Note 1 and 2 for band structure and quasi-2D approximation) [28,29].

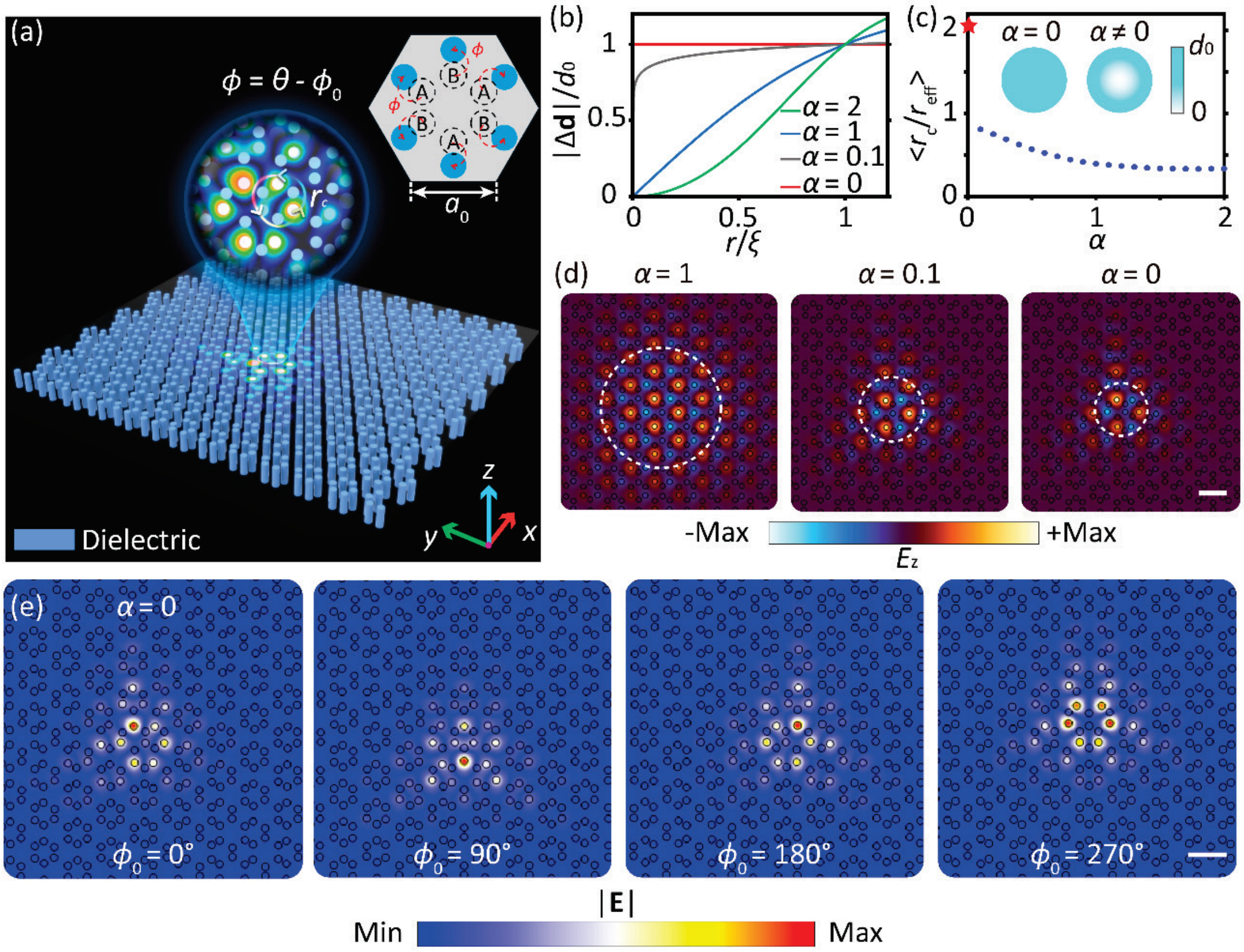

**FIG. 1. Initial phase-sensitivity of DVM after area contraction.** (a) Schematic of the quasi-2D topological PhC supporting a TM DVM induced by Kekulé modulation of the phase $\phi = \theta - \phi_0$, with $\theta$ being polar angle and $\phi_0$ being a global initial phase. The PhC is formed by dielectric rods arranged in a honeycomb lattice placed between parallel metallic plates. Rod radius $R = 2$ mm, lattice constant of primitive cell $a_0 = 12$ mm,

rod height $h$ = 10 mm. The magnified view highlights trajectory of energy-weighted center $\mathbf{r}_c$ of DVM driven by the initial phase $\phi_0$, with arrows indicating direction of increasing $\phi_0$. Inset: in-plane displacements $\Delta\mathbf{d}$ of A- and B-sublattices within an enlarged cell that realize the Kekulé modulation of phase $\phi$. (b) Radial profiles of the normalized Kekulé modulation strength for different shape factors $\alpha$. (c) Dimensionless ratio $\langle r_c/r_{\text{eff}}\rangle$ of the mode center offset $r_c$ to the effective mode radius $r_{\text{eff}}$ for different shape factors $\alpha$, with $\langle\ \rangle$ representing average over the initial phase $\phi_0$. Inset: schematics of the radial profiles for $\alpha = 0$ (highlighted by red star) and $\alpha \neq 0$ (blue dots). (d) Electric field maps of DVMs for different shape factors $\alpha$, with the white dashed circles outlining their characteristic areas. (e) Evolution of the DVM ($\alpha = 0$) driven by the varying global initial phase $\phi_0$. White scale bar: 20 mm.

To realize the DVM, we introduce a sublattice-resolved in-plane displacement of individual rods from their unperturbed positions. For the rod located at position $\mathbf{r}$, its displacement $\Delta\mathbf{d}$ is given by [inset in Fig. 1(a)]:

$$\Delta\mathbf{d} = d_0 \frac{\tanh[(r/\xi)^\alpha]}{\tanh(1)}\left[\sin\left(\mathbf{K}\cdot\mathbf{r}+\phi\right), \pm\cos\left(\mathbf{K}\cdot\mathbf{r}+\phi\right)\right] \tag{1}$$

where $\mathbf{K}$ = [$4\pi/(3a_0)$, 0] is the Kekulé vector, $d_0$ = $0.08\sqrt{3}a_0$ is the modulation strength and $\xi = 6\sqrt{3}a_0$ is the vortex radius we choose, with $a_0$ being the lattice constant of the primitive cell. Here, $\phi = \theta - \phi_0$ is the Kekulé modulation phase, with $\theta$ representing the azimuthal angle of the hexagonal enlarged cell, and the ± sign refers to the A and B sublattices, respectively. At this stage, the parameter $\phi_0$ manifests as the global initial phase of the Kekulé modulation, specifying the overall phase offset of the Kekulé bonding texture; later we will reveal that $\phi_0$ acts as a gauge DOF for the DVM in the continuum model. On the other hand, the radial distribution of this modulation is governed by the shape factor $\alpha$ [18], which gives a smooth ramp profile for non-zero $\alpha$ and transitions to a constant one for $\alpha = 0$ [Fig. 1(b)]. As $\alpha \to 0$, the DVM undergoes a significant area contraction, so that the offset of the mode from the geometric center of the PhC becomes significant. To quantify this point, we calculate the dimensionless ratio $\langle r_\mathrm{c}/r_\mathrm{eff} \rangle$, which characterizes the mode-center displacement $r_\mathrm{c}$ relative to the effective mode radius $r_\mathrm{eff}$, averaged over the initial phase $\phi_0$. Particularly, the offset $\mathbf{r}_\mathrm{c}$ is tracked by energy-density-weighted average of the mode:

$$\mathbf{r}_\mathrm{c} = \frac{\int \mathbf{r} \varepsilon_r |\mathbf{E}|^2 \, dA}{\int \varepsilon_r |\mathbf{E}|^2 \, dA}, \tag{2}$$

with $r_\mathrm{c} = |\mathbf{r}_\mathrm{c}|$, while the radius $r_\mathrm{eff}$ is extracted based on the effective mode area:

$$r_{\text{eff}} = \sqrt{\frac{1}{\pi}\frac{\int \varepsilon_r |\mathbf{E}|^2 \, dA}{\max\left(\varepsilon_r |\mathbf{E}|^2\right)}}. \tag{3}$$

The calculated ratio rises sharply when $\alpha = 0$, that is, the constant modulation regime [Fig. 1(c)]. In this regime, the mode center is no longer obscured by a broadly distributed field profile [Fig. 1(d)]. Instead, the relocation and redistribution of the DVM become directly visible in its field maps as $\phi_0$ varies [Fig. 1(e), see also SI Movie S1 for animation]. Consequently, the $\phi_0$-driven evolution of the mode results in a lattice-constant-scale motion of the DVM center relative to its highly contracted profile [magnified view in Fig. 1(a)]. A further hallmark of the constant-modulation regime is that the abrupt increase of the ratio $\langle r_{\text{c}}/r_{\text{eff}} \rangle$ essentially arises from the finite modulation amplitude $d_{\text{c}}$ (= $d_0$) of the central unit cell (CUC) at the geometric origin. Since the azimuthal angle $\theta$ is undefined at the origin, we explicitly assign a local Kekulé phase $\phi_{\text{c}}$ to the CUC that is fixed when $\phi_0$ varies. While the global initial phase $\phi_0$ drives a rotation-like motion of the DVM, the CUC with its Kekulé modulation amplitude $d_{\text{c}}$ and phase $\phi_{\text{c}}$ acts as a lever that dictates the radius of this motion (see SI Note 3 and Note 8).

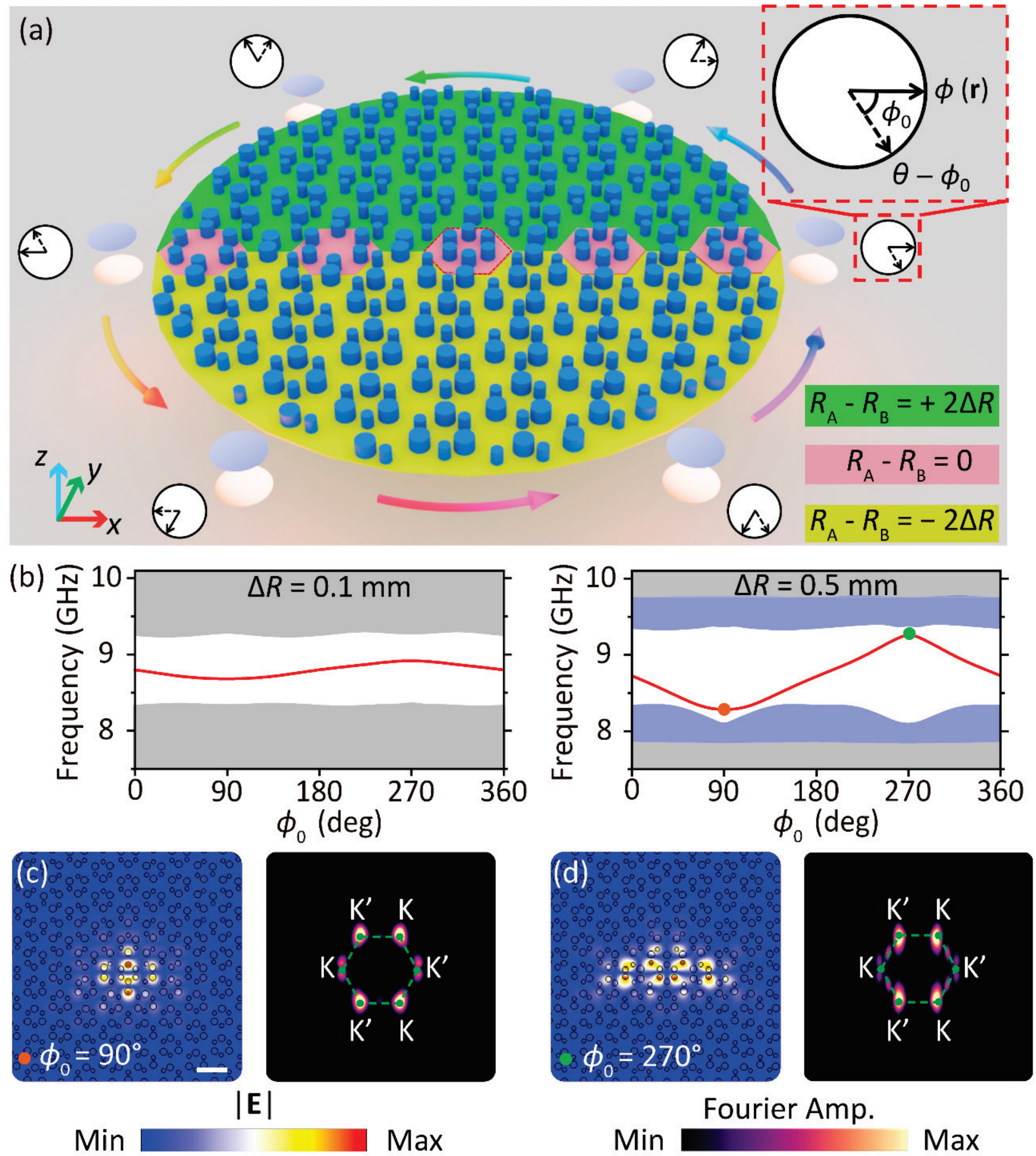

**FIG. 2. Initial phase-controlled frequency tuning of the DVM.** (a) Schematic representation of the PhC metastructure composed of three domains. Inset: the solid arrow represents the Kekulé phase and the dashed arrow represents $\theta - \phi_0$. The red dashed hexagon highlights the central unit cell (CUC). (b) Numerically computed eigenfrequency spectra of the metastructure as a function of $\phi_0$ for $\Delta R = 0.1$ mm and $\Delta R = 0.5$ mm. The blue (gray) shaded regions represent edge-like (bulk) modes. (c), (d) Simulated electric field maps (|**E**|) of the DVM for (c) $\phi_0$ = 90° and (d) 270°, showing strongly localized mode distributions. White scale bar: 20 mm. The corresponding spatial fast Fourier transform (FFT) amplitude spectra show pronounced spots around the K and K′ points. Green dashed hexagons: first Brillouin zone (FBZ) of the primitive cell.

In the following, we demonstrate a direct consequence and application of this mode-center motion controlled by the initial phase $\phi_0$. We fix the CUC modulation phase $\phi_c = 0°$ to maximize the mode-center motion. To translate this initial phase-driven spatial motion into a continuous frequency shift, we introduce a sublattice-antisymmetric perturbation across the mirror symmetry axis ($y = 0$) as shown in Fig. 2(a). This configuration effectively exposes the DVM to regions with contrasting sublattice biases. The spatial profile of the radius difference can be written as:

$$R_A - R_B = 2\Delta R \left|\tanh\left(\frac{y}{\xi}\right)\right|^{\beta} \mathrm{sgn}(y)\,. \tag{4}$$

Here, $\beta$ controls the sharpness of the transition between the two sublattice-biased regions. In practice, we choose $\beta = 0.01$ to optimize the frequency tuning range. Results for smoother transitions, such as $\beta = 0.5$, are given in SI Note 4. This design can effectively suppress the leakage of DVM caused by strong scattering due to abrupt structural discontinuities. As $\phi_0$ varies, the driven mode center drags the DVM to traverse between these two regions, allowing the mode to dynamically sample distinct sublattice biases. We first consider the case of a small perturbation $\Delta R = 0.1$ mm, for which the DVM exhibits a frequency shift that varies with $\phi_0$ in a sinusoidal-like manner [Fig. 2(b)]. When the perturbation is increased to $\Delta R = 0.5$ mm, the enhanced sublattice contrast between the regions substantially amplifies the frequency shift, resulting in a frequency tuning range that nearly spans the entire original topological bandgap [Fig. 2(b)]. See SI Movie S2 for the $\phi_0$-driven evolution of the PhC and corresponding spectral response.

The corresponding field maps at the minimum and maximum frequencies, that is, when $\phi_0 = 90°$ and 270° are shown [Fig. 2(c), (d)]. In both cases, the mode remains highly localized around the vortex core. In the accompanying Fourier spectra, the clear signatures of the bright spots around the K and K′ points also confirm that the mode retains a coherent superposition of the two valley components during the $\phi_0$-driven spatial motion.

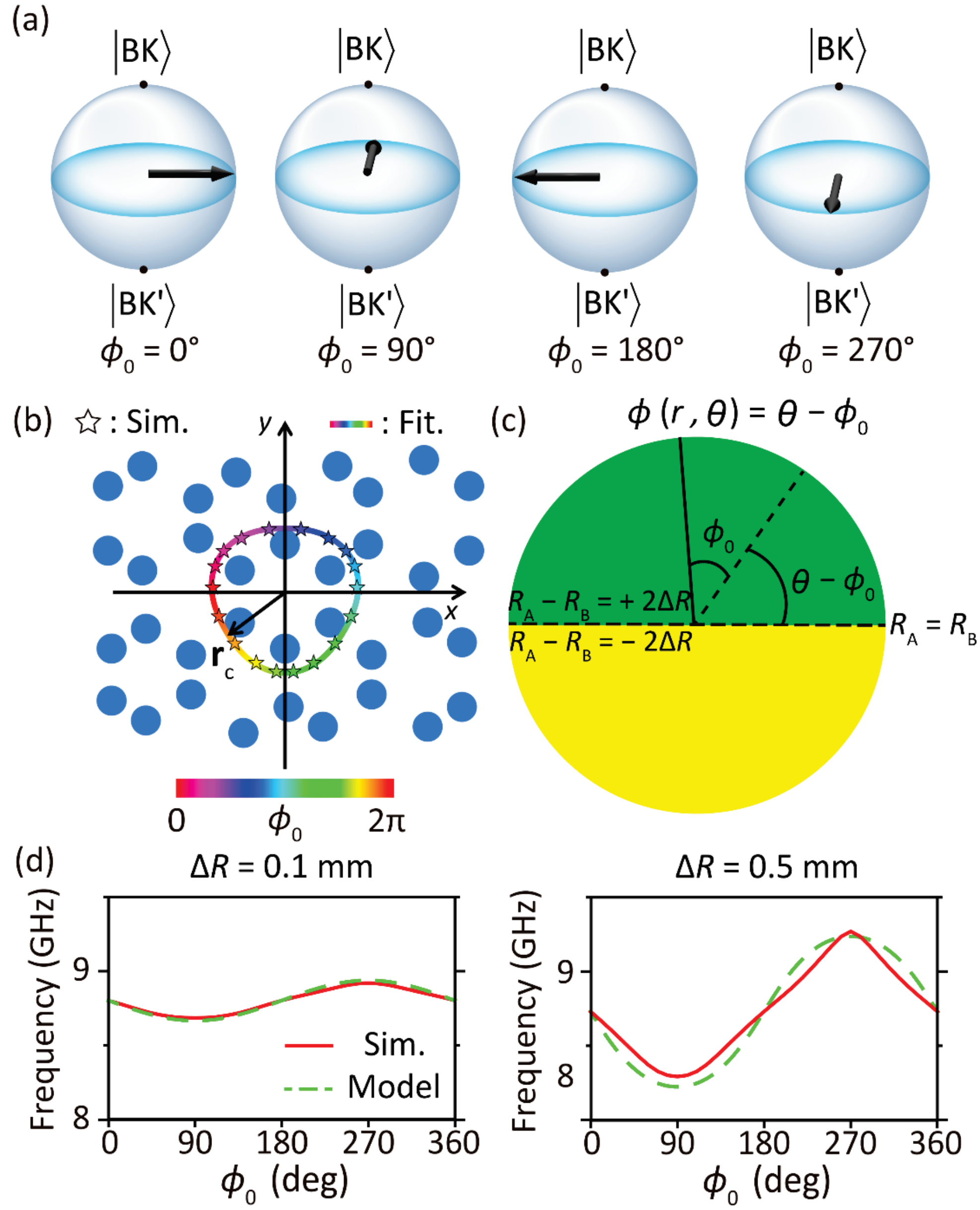

**FIG. 3. Effective Hamiltonian modeling of initial phase sensitivity.** (a) DVM in the continuum limit represented on the valley Bloch sphere with $|\mathrm{BK}\rangle$ and $|\mathrm{BK'}\rangle$ being the poles and $\phi_0$ acting as the azimuthal angle. (b) Trajectory of the energy-density-weighted DVM's centers ($\mathbf{r}_c$) as a function of $\phi_0$. The markers are directly extracted from full-wave simulations with different values of $\phi_0$, and the superimposed colored line is fitted to them.

(c) A continuum model of the PhC in Fig. 2(a). (d) Comparisons between the semi-analytical calculations (from the perturbative model) and full-wave simulations of the frequency spectra of the DVM as $\phi_0$ varies, for $\Delta R$ = 0.1 mm and $\Delta R$ = 0.5 mm.

To analyze impact of the initial phase $\phi_0$, we start from the standard effective Hamiltonian for Kekulé modulation when there is no radius difference:

$$H = v_{\mathrm{D}}(k_x\sigma_x \otimes \tau_z + k_y\sigma_y \otimes I) + m(\mathbf{r})\sigma_x \otimes [\tau_x \cos\phi + \tau_y \sin\phi], \tag{5}$$

where $\boldsymbol{k} = (k_x, k_y)$ is the Bloch wavevector, $v_{\mathrm{D}}$ is the Dirac velocity near the Γ point, $\sigma_i$ and $\tau_i$ (for $i = x, y, z$) are Pauli matrices acting on the sublattice and valley pseudospins, respectively, and $I$ is the identity matrix. Mapping the discrete in-plane displacements $\Delta\mathbf{d}$ defined in Eq. (1) to the complex modulation mass, we have:

$$m(\mathbf{r}) = m_0 \frac{\tanh[(r/\xi)^{\alpha}]}{\tanh(1)}, \tag{6}$$

where $m_0$ is controlled by $d_0$. With the substitution ($k_x \to -i\partial_x$, $k_y \to -i\partial_y$), the effective Hamiltonian supports a localized zero mode at the Dirac frequency, which in the basis of $\{|\mathrm{AK}\rangle, |\mathrm{AK'}\rangle, |\mathrm{BK}\rangle, |\mathrm{BK'}\rangle\}$ can be written as [30,31]:

$$\Psi_0 = \mathcal{N} e^{-\frac{m_0}{v_D \tanh(1)} \int_0^r \tanh(r'/\xi)^\alpha dr'} \begin{pmatrix} 0 \\ 0 \\ e^{i\phi_0/2} \\ -ie^{-i\phi_0/2} \end{pmatrix}. \tag{7}$$

Here, $\mathcal{N}$ is the normalization constant. The spinor solution shows that the DVM is strictly polarized on the B sublattice (see SI Note 5), forming an equal-amplitude superposition of the K and K' valleys. In this continuum model, $\phi_0$ is a gauge DOF acting as an azimuthal angle denoting the intervalley phase between $|\mathrm{BK}\rangle$ and $|\mathrm{BK'}\rangle$, which can be absorbed by a U(1) rotation on the valley Bloch sphere [Fig. 3(a)]. However, this initial phase is no longer redundant in a discrete lattice, because changing $\phi_0$ shifts the Kekulé bonding texture around the vortex core which directly affects the mode distribution [32], leading to a measurable displacement of the DVM. To account for this correction, we introduce the displaced mode center $\mathbf{r}_c$ into the ideal wave function (see SI Note 5):

$$\Psi'_0(\mathbf{r},\phi_0) = \Psi_0(\mathbf{r}-\mathbf{r}_c,\phi_0). \tag{8}$$

Without this displacement correction, the ideal continuum wavefunction remains centered at the vortex core and the first-order expectation value of the antisymmetric perturbation vanishes, giving no phase-dependent frequency shift. As mentioned previously (Fig. 1), the displacement can be extracted from simulation results by taking the energy-weighted averaging of the DVM over varying $\phi_0$ that traces out a trajectory [Fig. 3(b)]. In addition, the antisymmetric perturbation partitions the continuum model into two domains separated by the $x$-axis [Fig. 3(c)], while the corresponding additional sublattice-staggered mass term can be modeled as:

$$H_\Delta(\mathbf{r}) = -m_\Delta\left[\left|\tanh(\frac{y}{\xi})\right|^\beta \operatorname{sgn}(y)\right](\sigma_z \otimes I), \tag{9}$$

where $m_\Delta$ is controlled by $\Delta R$. Treating this term as a first-order perturbation, we consider the ideal limit as shown in Fig. 3(c), where $\beta \rightarrow 0$. The DVM frequency shift can be obtained from the expectation value (see SI Note 5):

$$\begin{aligned}\Delta f^{(1)} &= \langle \Psi_0' | H_\Delta | \Psi_0' \rangle \\ &= -m_\Delta \operatorname{sgn}(\sin\phi_0)\frac{2}{\pi}\int_0^{2\frac{m_0}{v_D}r_0|\sin\phi_0|} tK_1(t)dt\end{aligned}, \tag{10}$$

where $K_1(t)$ is the modified Bessel function of the second kind, and $r_0$ is the radius of the circle fitted to the trajectory. With all the parameters extracted from full-wave simulations, Eq. (10) explicitly reveals that the frequency shift arises from the mode-center displacement $\mathbf{r}_c$, indicating that without the correction introduced in Eq. (8), the first-order expectation value cancels by symmetry and no initial phase-dependent spectral response appears. Meanwhile, by comparing the semi-analytic prediction with the simulation results when $\Delta R$ = 0.1 mm [Fig. 3(d)], we observe excellent agreement as the frequency varies sinusoidally with $\phi_0$. This sinusoidal behavior can be explained from the sin$\phi_0$ dependence of the dominant first term after expansion of Eq. (10), which directly results from the rotation-like motion of the DVM center driven by $\phi_0$ (see SI Note 5). Intuitively, when $\phi_0 = 0°$ or 180°, the DVM center lies on the $x$-axis and the contributions from the two domains cancel each other, giving $\Delta f^{(1)} = 0$. As $\phi_0$ deviates, the imbalance between the two domains grows and $|\Delta f^{(1)}|$ increases. When $\Delta R$ increases to 0.5 mm, both analysis and simulation results still exhibit a sinusoidal-like dependence on $\phi_0$, although higher-order terms distort the spectrum to some extent [Fig. 3(d)]. Furthermore, we provide other examples to validate the semi-analytical effective Hamiltonian model (see SI Note 6 and 7). In general, the effective Hamiltonian model captures the essential valley physics and sinusoidal-like behavior of our approach for continuously controlling the DVM's frequency through the initial phase $\phi_0$.

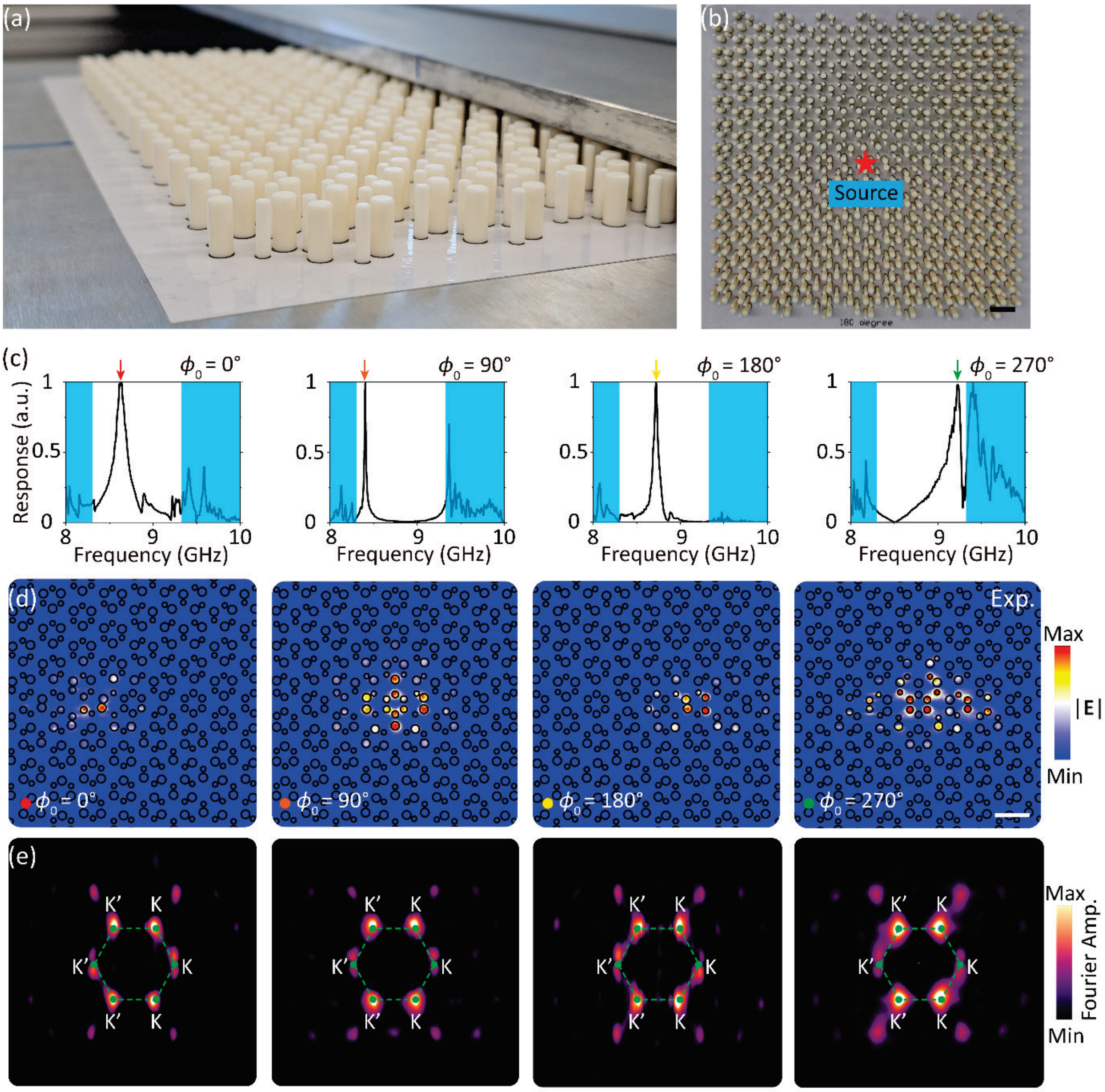

**FIG. 4. Experimental demonstration of the initial phase dependence.** (a) Perspective view of the fabricated PhC sample consisting of dielectric rods sandwiched between two parallel metallic plates to ensure a quasi-2D environment. (b) Top-view photograph of a sample with $\phi_0 = 180°$ (metallic cover removed). The red star indicates the location of the near-field excitation source. Black scale bar: 20 mm. (c) Measured response spectra of the samples with different initial phases, $\phi_0$= 0°, 90°, 180°, and 270°,

respectively. Colored arrows denote the peaks corresponding to the DVMs. The blue regions are the spectral ranges outside the bandgap. (d) Experimentally imaged electric field maps (|**E**|) at the frequency of corresponding peaks in (c), showing strongly localized mode distributions. White scale bar: 20 mm. (e) FFT amplitude spectra of the corresponding field maps in (d), revealing simultaneous occupation of both valleys. Green dashed hexagons: FBZ of the primitive cell.

To experimentally validate our findings, we constructed samples of the PhC metastructure, consisting of dielectric rods arranged between two metallic plates [Fig. 4(a)], with an excitation source inserted in the center [Fig. 4(b)] [33–36]. The electric field response was detected with a monopole antenna connected to a vector network analyzer (VNA) that recorded the $S_{21}$ parameter. In the measured frequency responses [Fig. 4(c)] of the samples for $\phi_0$ = 0°, 90°, 180°, and 270°, a clear resonance emerges within the bandgap. Moreover, as $\phi_0$ evolves through 360°, the resonance frequency undergoes a sinusoidal-like oscillation between the lower and upper edges of the bandgap, in good agreement with both analytical and numerical results.

We further mapped the electric field distributions inside the metastructure through the point-wise scanning of $S_{21}$ by moving the mounted sample. The measured electric field maps [Fig. 4(d)] at the resonance frequency for $\phi_0$ = 0°, 90°, 180°, and 270° all exhibit strong localization near the vortex core of the metastructure, consistent with the simulations in Fig. 2. The frequency offset is mainly due to unavoidable, inhomogeneous submillimeter air gaps between the dielectric rods and the metallic plate [27], which alter the effective permittivity, thereby slightly shifting the resonance frequency (see SI Note 2). Furthermore, we perform spatial FFTs to examine the resonances in the reciprocal space. The experimental Fourier spectra [Fig. 4(e)] of the field maps confirm the simultaneous occupation of the K and K′ valleys by these excited resonances. The strong consistency between experiment and simulation provides direct evidence for the initial phase-driven tunability of the DVM.

**Discussion**

In summary, we have shown that the initial phase of the Kekulé modulation, which is redundant in the continuum Jackiw–Rossi model, becomes physically observable in discrete lattices by sensitively driving a displacement of the DVM center. While the Jackiw–Rossi model correctly captures the topological origin and zero-mode nature of DVMs realized in discrete lattices, it smooths out the texture structure of the vortex core that is essential for resolving the effects of different initial phases. For a strongly localized DVM in the constant-amplitude modulation regime, the choice of initial phase produces a displacement comparable to the effective mode radius. By further introducing a sublattice-antisymmetric perturbation, this displacement can be converted into a sinusoidal-like spectral response spanning nearly the entire bandgap, a mechanism that cannot be captured within the conventional continuum description.

Meanwhile, by incorporating the lattice-induced displacement of the mode center into the continuum wavefunction within a perturbative framework, we obtain a correction that quantitatively reproduces the observed spectral response with respect to the initial phase. Therefore, our results highlight the breakdown of the ideal Jackiw–Rossi model and underscore the importance of a lattice-resolved paradigm for the study of DVMs, especially the strongly localized ones. In this sense, the initial phase is elevated from a redundant gauge choice to a genuine control and design parameter for reconfigurable photonic devices. More broadly, our work reveals that lattice discreteness could provide active DOFs for engineering topological phenomena in wave systems beyond the continuum limit.

## Methods

In experimental measurements, the photonic crystal (PhC) metastructure was fabricated using alumina rods with a nominal purity of $Al_2O_3$ > 99% and a relative permittivity of approximately 9.8. The rods were arranged between two parallel metallic plates to construct the experimental samples. Microwave measurements were performed using two probes connected to a VNA (Ceyear 3674H). The source is connected to port 1 of the VNA, while the detector is connected to port 2 of the VNA.

In numerical simulations, we used COMSOL Multiphysics. For band structure calculations, two-dimensional and three-dimensional simulations were carried out for both primitive-cell and enlarged-cell models. Floquet periodic boundary conditions were applied along the hexagonal directions to account for the in-plane periodicity. In the three-dimensional enlarged-cell simulations, perfect electric conductor (PEC) boundary conditions were applied along the z direction to model the metallic plates used in the experiments. For finite-structure simulations, scattering boundary conditions (SBCs) were applied to the outer boundaries to approximate an open electromagnetic environment.

## Data availability

The data which support the figures and other findings within this paper are available from the corresponding authors upon request.

## Author Contributions

Jiayu Fan, Haitao Li and Xiaoxiao Wu designed research; Jiayu Fan, Jiusi Yu, Aoning Luo, and Yiyi Yao performed research; Jiayu Fan, Shijie Kang, Xiexuan Zhang, and Haitao Li analyzed data; Jiayu Fan and Haitao Li prepared the visualization; Jiayu Fan and Xiaoxiao Wu wrote the paper; Biye Xie, Xiao-Dong Chen, and Xiaoxiao Wu reviewed and edited the manuscript.

## Acknowledgement

This research was supported by the National Natural Science Foundation of China (No. 12304348, 62475225, 12404187, 12522415), Guangdong Basic and Applied Basic Research Foundation (No. 2025A1515011470, 2024A1515012031), Guangdong University Featured Innovation Program Project (2024KTSCX036), Guangzhou-HKUST(GZ) Joint Funding Program (2025A03J3783), HKUST — HKUST(GZ) 20 for 20 Cross-campus Collaborative Research Scheme (G011), National Key R&D Program of China (No. 2023YFA1407700), Shenzhen Science and Technology Innovation Commission (No. JCYJ20240813113619025).

## Competing interests

The authors declare no competing financial interests.